\documentclass[twocolumn,aps,showpacs,prl,amsmath,amssymb,floatfix,superscriptaddress]{revtex4}
\usepackage{graphicx}
\usepackage{dcolumn}
\usepackage{bm}
\usepackage{color} 
\usepackage{array}
\usepackage{float}
\usepackage{supertabular}
\usepackage{longtable}

\begin{document}

\title{Wealth Distributions in Asset Exchange Models}

\author{P.~L.~Krapivsky and S. Redner}
\affiliation{Department of Physics, Boston University, Boston, MA 02215}

\begin{abstract}

  How do individuals accumulate wealth as they interact economically?  We
  outline the consequences of a simple microscopic model in which repeated
  pairwise exchanges of assets between individuals build the wealth
  distribution of a population.  This distribution is determined for generic
  exchange rules --- transactions that involve a fixed amount or a fixed
  fraction of individual wealth, as well as random or greedy exchanges.  In
  greedy multiplicative exchange, a continuously evolving power law wealth
  distribution arises, a feature that qualitatively mimics empirical
  observations.
\end{abstract}


\maketitle

\section{PERSPECTIVE}

The economy is a complex interacting system that responds to a multitude of
influences and extends over a wide range of monetary scales.  As experience
with financial crises continues to demonstrate, understanding how an economy
develops and how it is influenced by externalities remains poorly understood.
Basic questions about what causes financial crises and how to deal with them
continue to be hotly debated, with little sign that a fundamental
understanding is emerging~\cite{A88,B88,K93}.

What can statistical physics contribute to this discussion?  Not much, if the
goal is to predict the economy next year.  However, statistical physics
possesses powerful theoretical tools that have proven useful in describing
specific financial phenomena, such as the Black-Scholes options pricing
formula~\cite{BS73}.  There are many parallels between statistical physics
and economic phenomena, and physics-based modeling has helped facilitate
conceptual developments in finance and economics (see
e.g.,~\cite{BP00,MS00,F05}).

In classic economic theories, humans, or companies, are considered as
rational actors that respond deterministically to external conditions.  More
recently, stochastic tools have been applied to the economy, particularly to
financial modeling.  The stochastic approaches that are conventionally
employed are Brownian motion and its generalizations.  In physics, a similar
approach was followed to describe non-deterministic systems, where the
interaction between a particle and its environment was mimicked by noise,
while interactions between microscopic entities (such as Brownian particles)
were ignored.  This development (associated with physicists like Einstein,
Langevin, and Stratonovich, and mathematicians like Kolmogorov, Feller, and
It\^o) led to increasingly sophisticated stochastic processes
\cite{chandra,ito,feller,D06}, a research thread that is still active.

Over the last 40 years a new approach that combines the stochastic behavior
of elemental entities as well as their mutual interactions has emerged (see
e.g.,~\cite{L99,KRB10}).  While a few-particle interacting system is
hopelessly complicated and beyond the reach of analytical techniques, a
dramatic simplification arises for a many-particle system because we can
often make {\em statistical} predictions about its fate.  That a macroscopic
interacting system is simpler than its few-element counterpart has several
appellations --- the law of large numbers, ergodicity, etc. --- and it
justifies the utility of statistical physics and probability theory in
attempts to understand economic processes.

\begin{figure}
\begin{center}
\includegraphics[width=0.35\textwidth]{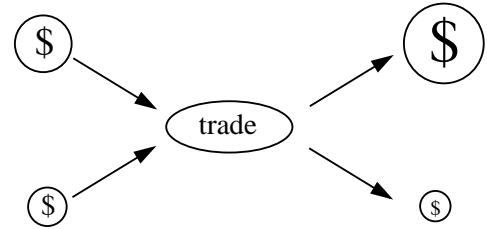}
\caption{Illustration of the asset exchange model}
\label{trade}
\end{center}
\end{figure}

In this short review, we present an interacting many-agent {\em asset
  exchange} model that can be quantitatively analyzed using statistical
physics tools.  An agent could be a single person or a self-contained
economic entity, such as a company.  In this model, the interaction between
two agents results in a redistribution of their assets.  We regard an asset
as any economic attribute --- cash, goods, or other materials --- that
contributes to overall individual wealth.

The macroeconomy is viewed as the result of a large number of asset exchanges
between randomly-selected pairs of agents.  Through these exchanges a global
wealth distribution develops, and we want to understand how generic features
of this distribution depend on the nature of the exchanges.  The notion that
the wealth distribution is driven by two-person exchanges appears to have
been first considered in the economics literature by Angle~\cite{A1,A2}.  In
the physics community, this approach was introduced by Ispolatov et
al.~\cite{IKR98}, and related perspectives on this subject include
Refs.~\cite{DY00,BM00,CC00,S04,CC07}.  A comprehensive review of this
research topic is given in~\cite{YR09} and an engaging non-technical
exposition appears in~\cite{H02}.

\section{ADDITIVE EXCHANGE}

As a preliminary, we first study additive exchange processes, in which a
fixed amount of asset is exchanged between two agents, independent of their
wealth before a trade occurs.  At the outset, we have to determine how to
treat agents whose wealth reaches zero as a result of many unfavorable
trades.  We treat such penurious agents as economically ``dead'', so that
they no longer participate in the evolution of the wealth.  Mathematically,
this rule corresponds to imposing an absorbing boundary condition on the
density of agents of zero wealth.  An alternative is to impose a reflecting
boundary condition at zero wealth, so that all agents continue to
economically interact, even if they have no wealth~\cite{DY00}.  In this
latter case, the wealth follows the Boltzmann distribution of equilibrium
statistical mechanics, with an effective temperature equal to the average
amount of wealth per agent.

Under the condition that bankrupt agents are eliminated from further economic
activity, we determine the consequences of: (i) {\em fair} transactions,
where either agent is equally likely to profit in an interaction, and (ii)
{\em greedy} transactions, in which the richer agent profits in an
interaction.  For simplicity, each agent is assumed to possess an
integer-valued amount of assets and that one unit of asset is transferred
between traders in each interaction.

\subsection{Fair Transactions}

In a fair exchange, the wealth of two agents evolves as $(j,k)\to (j\pm
1,k\mp 1)$; the direction of the exchange is independent of their starting
wealth.  The wealth distribution evolves by selecting two agents at random
who exchange one unit of wealth and repeating this elemental step {\em ad
  infinitum}.  We assume that all agents are equally likely to interact with
any other agent (corresponding to the mean-field limit in statistical
physics).

In this limit, the evolution of the wealth distribution is described by a
{\em master equation} that accounts for the changes in wealth in each
microscopic interaction between agents.  Let $c_k(t)$ be the density of
agents with wealth $k$.  In random additive exchange, the master equation is
\begin{equation}
\label{ckran}
\frac{dc_k}{dt}=N\left[c_{k+1}+c_{k-1}-2c_k\right],
\end{equation}
where $N(t)\equiv \sum_{k\geq 1} c_k(t)$ is the density of economically
viable agents.  The first two terms on the right-hand side account for the
gain in $c_k$ due to the transactions $(j,k+1)\to (j+1,k)$ and $(j,k-1)\to
(j-1,k)$, respectively, while the last term accounts for the loss in $c_k$
due to the transactions $(j,k)\to (j\pm 1,k\mp 1)$.  Since these transactions
require the presence of an agent of wealth $k$ or $k\pm 1$ and an agent of
arbitrary wealth, all terms on the right-hand side involve $N$ times a
concentration.  The density of agents with a single unit of wealth evolves by
$dc_1/dt=N(c_2-2c_1)$; this equation may also be written in the same form as
Eq.~(\ref{ckran}) by imposing the absorbing boundary condition $c_0(T)=0$.

Introducing the time-like variable, $T=\int_0^t dt' N(t')$, we reduce
Eq.~(\ref{ckran}) to the discrete diffusion equation
\begin{equation}
\label{diff}
\frac{dc_k}{dT}=c_{k+1}+c_{k-1}-2c_k\,,
\end{equation}
which may be solved for any initial condition~\cite{feller,KRB10}.  When all
agents start with unit wealth, $c_k(0)=\delta_{k,1}$, we may account for the
absorbing boundary condition by augmenting the initial condition with an
``image'' contribution due to agents with initial wealth $-1$; that is,
$c_k(0)=\delta_{k,1}-\delta_{k,-1}$.  The solution to Eq.~(\ref{diff})
subject to these initial conditions is~\cite{KRB10}
\begin{eqnarray}
\label{cksol}
c_k(T)=e^{-2T}\left[I_{k-1}(2T)-I_{k+1}(2T)\right]\,,
\end{eqnarray}
where $I_n$ is the modified Bessel function of order $n$.  Correspondingly,
the total density of active agents $N(T)$ is
\begin{eqnarray}
\label{Nsol}
N(T)=e^{-2T}\left[I_0(2T)+I_1(2T)\right]\,.
\end{eqnarray}

In the limit $T\to\infty$, the asymptotic behaviors of Eqs.~\eqref{cksol} and
\eqref{Nsol} are:
\begin{equation}
\label{ckN-asymp}
c_k\simeq \frac{k}{\sqrt{4\pi T^3}}\,\, e^{-k^2/4T}\,, \qquad\qquad
N \simeq (\pi T)^{-1/2}~.
\end{equation}
These asymptotics apply, up to an overall factor, to all initial conditions
that decay sufficiently rapidly with $k$.  An important feature of
\eqref{ckN-asymp} is the emergence of {\em scaling\/}: the distribution
$c_k(T)$ depends on the scaled wealth, $k/\sqrt{T}$, rather than separately
on the variables $k$ and $T$.  Similar scaling behavior arises in numerous
interacting particle systems~\cite{KRB10}.  Normally, scaling is {\em
  postulated\/} and then verified analytically or numerically.  For asset
exchange, we deduce the validity of scaling from the exact solution.  We now
express the asymptotic solution \eqref{ckN-asymp} in terms of the physical
time $t$ by using $t(T)=\int_0^T dT'/N(T')\simeq \frac{2}{3}\sqrt{\pi T^3}$
to eliminate $T$ and give
\begin{equation}
c_k\simeq \frac{k}{3t}\,
\exp\left[-\left(\frac{\pi}{144}\right)^{1/3}\,\frac{k^2}{t^{2/3}}\right],\qquad
N\simeq \left(\frac{2}{3\pi t}\right)^{1/3}
\end{equation}
The number of viable agents decreases as $t^{-1/3}$ and their typical wealth
grows as $t^{1/3}$.  While this model is not realistic, it illustrates the
efficacy of a statistical physics perspective in solving an interacting
many-body system.

Instead of removing bankrupt agents, let us provide each of them with
`welfare' of a single unit of asset.  In this case, the economically viable
population density is always $N=1$ and the master equation for the wealth
distribution simplifies to
\begin{equation}
\begin{split}
\label{ckrandom}
\frac{dc_k}{dt}=c_{k+1}+c_{k-1}-2c_k\,\qquad k\geq 2\,,\\
\frac{dc_1}{dt}=c_2-c_1\,, ~\qquad \qquad \qquad k=1\,.
\end{split}
\end{equation}
We can extend the first of these equations to all $k$ and also subsume the
equation for $c_1$ by choosing the initial condition $c_{1-k}(0)=c_k(0)$,
with $c_1(0)=c_0(0)=1$, and $c_k(0)=0$ for $k\ne 0, 1$. The solution to
\eqref{ckrandom} subject to this initial condition is
\begin{eqnarray}
\label{ckt-sol}
c_k=e^{-2t}\left[I_{k-1}(2t)+I_{k}(2t)\right]\,.
\end{eqnarray}
Because of this injection of assets to destitute agents, the total wealth
density of the population, $M=\sum_{k\geq 1}kc_k$, grows with time as
\begin{eqnarray*}
M = e^{-2t}\sum_{k\geq 1}k\left[I_{k-1}(2t)+I_{k}(2t)\right]
    \simeq 2\sqrt{t/\pi}\,,
\end{eqnarray*}
as $t\to\infty$.  In this toy model, the rate of welfare expenditure to keep
everyone solvent decreases with time!

\subsection{Greedy Transactions}

In greedy exchange, the richer agent is exploitative and always takes one
unit of wealth from the poorer agent in each interaction, as represented by
$(j,k)\to (j+1,k-1)$ for $j\geq k$.  The densities $c_k(t)$ now evolve
according to
\begin{equation}
\label{ck}
\frac{dc_k}{dt}=c_{k-1}\sum_{j=1}^{k-1}c_j+c_{k+1}\sum_{j=k+1}^\infty c_j
-c_k( N+c_k).
\end{equation}
The first term on the right accounts for the gain in $c_k$ due to an agent
with wealth $k-1$ taking one wealth unit from a poorer trading partner.
Similarly, the second term accounts for an agent with wealth $k+1$ losing one
unit of wealth to a richer trading partner.  The last term accounts for the
loss of $c_k$ when an agent of wealth $k$ trades with anyone; the extra
factor of $c_k$ accounts for the loss of both agents of wealth $k$ when two
such agents interact.

While this set of non-linear equations appears intractable by exact methods,
they are readily amenable to a scaling analysis~\cite{KRB10}.  We first
re-write Eq.~(\ref{ck}) as
\begin{equation}
\begin{split}
\label{ckt}
\frac{dc_k}{dt}=-c_k(c_k+c_{k+1})+N(c_{k-1}-c_k)~~~~~~~~~~~~~\\
               ~~~~~~+(c_{k+1}-c_{k-1})\sum_{j=k}^\infty c_j,
\end{split}
\end{equation}
and make the {\em scaling ansatz} $c_k(t)\simeq
k_*^{-2}\,\mathcal{C}(k/k_*)$, where $k_*(t)$ is the typical wealth of each
agent.  That is, the wealth distribution at different times is invariant when
wealth is measured in units of the time-dependent typical wealth.  The
prefactor $k_*^{-2}$ ensures that the total wealth of the population, $\sum_k
k c_k(t)$, is conserved, while the condition $\sum_k c_k(t)=N(t)$ gives
$k_*(t)\sim 1/N(t)$.  Substituting now the scaling form $c_k(t)\simeq
N^2\,\mathcal{C}(x)$, with $x=kN$, in Eq.~(\ref{ckt}) and taking the
continuum limit gives
\begin{equation}
\label{cx}
\mathcal{C}(0)[2\mathcal{C}+x\mathcal{C}']=2\mathcal{C}^2
+\mathcal{C}'\Big[1-2\int_x^\infty dy\,\mathcal{C}(y)\Big],
\end{equation}
where $\mathcal{C}'=d\mathcal{C}/dx$.  The scaling function must satisfy
\begin{equation}
\label{rel}
\int_0^\infty dx\,\,\mathcal{C}(x)=1, \quad{\rm and}\qquad
\int_0^\infty dx\,\,x\,\,\mathcal{C}(x)=1,
\end{equation}
that follow from $N=\sum c_k(t)\simeq N\int dx\,\mathcal{C}(x)$ and
setting the (conserved) wealth density to one, $\sum_k kc_k(t)=1$.

Equation~(\ref{cx}) is soluble by elementary techniques~\cite{IKR98}, and the
asymptotic wealth is simply the step function
\begin{equation}
\label{fermi}
c_k(t)=
\begin{cases}
1/(2t),       &k<2\sqrt{t}\,,\\
     0,            &k\geq 2\sqrt{t}\,,
\end{cases}
\end{equation}
while the density of active agents decays as $N(t)=t^{-1/2}$.  In greedy
exchange, the number of viable agents decays faster than in random exchange
and the population is slightly wealthier, with the average wealth growing as
$t^{1/2}$ rather than as $t^{1/3}$.

\section{MULTIPLICATIVE EXCHANGE}

While additive exchange provides instructive warm-up examples, multiplicative
exchanges, where a fixed fraction of the current wealth of one of the agents
is traded, are economically more realistic.  For example, investment returns
are generally quoted as percentages rather than absolute amounts.  A trade
now has the form $(x,y)\to (x-\alpha x, y+\alpha x)$, with $0<\alpha<1$ the
fraction of the loser's assets that are gained by the winner.  By
multiplicative exchanges agents can never go bankrupt, but they can become
arbitrarily poor.

\subsection{Fair Transactions}

If an agent gains or loses with equal probabilities in a transaction, the
wealth distribution evolves as
\begin{eqnarray}
\label{consint}
\frac{\partial c(x)}{\partial t}&\!\!=\frac{1}{2}
\int\!\!\int dy\,dz\,c(y)c(z)\!\times\! \big[-\delta(x-z)-\delta(x-y)\nonumber\\
&+\delta(y(1-\alpha)-x)+\delta(z+\alpha y-x)\big]\,.
\end{eqnarray}
The delta functions cleanly indicate the origin of the various terms in this
equation.  For example, the first two terms on the right account for the loss
of agents of wealth $x$ due to trades with any other agents.  The next two
terms account, respectively, for the gain in $c(x)$ due to the exchanges
$(\tfrac{x}{1-\alpha},y)\to(x,y+\tfrac{\alpha x}{1-\alpha})$ and $(y,x-\alpha
y)\to(y(1-\alpha),x)$.  Integrating over the delta functions, the master
equation becomes
\begin{equation}
\begin{split}
\label{cons}
\frac{\partial c(x)}{\partial t}= -c(x)+\tfrac{1}{2(1-\alpha)}\,\,
c\big(\tfrac{x}{1-\alpha}\big)~~~~~~~~~~~~\\
~~~~~~+\tfrac{1}{2\alpha}\int_0^x dy\,c(y)\,c\big(\tfrac{x-y}{\alpha}\big),
\end{split}
\end{equation}
where we set the (conserved) total density to one.

Equation~\eqref{cons} is daunting, and it is simpler to study the evolution
of the moments, $M_n(t)\equiv\int_0^\infty dx\,x^n \,c(x,t)$, that quantify
the wealth of a typical agent.  It is straightforward to verify that the
first two moments, the population $M_0$ and the wealth density $M_1$, are
conserved; we choose $M_0=1$ and $M_1=M$ without loss of generality.  More
interesting behavior arises for the second moment equation
\begin{equation}
\label{mom2}
\frac{d M_2(t)}{dt}=-\alpha(1-\alpha)M_2(t)+\alpha M^2~,
\nonumber
\end{equation}
whose solution is
\begin{equation}
\label{m2}
M_2(t)=\frac{M^2}{1-\alpha}
+\left[M_2(0)-\frac{M^2}{1-\alpha}\right]\,e^{-\alpha(1-\alpha)t}~.
\end{equation}
All moments beyond the second also converge to non-zero steady-state values.
This steady state arises because the wealth of a rich agent substantially
diminishes in a losing multiplicative exchange, but its wealth increases only
slightly in a winning exchange.  Conversely, a poor agent suffers a slight
loss in a losing exchange but can gain substantially in a winning exchange.
These two countering outcomes tends to move all agents toward a middle class.

\subsection{Greedy Transactions}

\begin{figure}
\begin{center}
\includegraphics[width=0.4\textwidth]{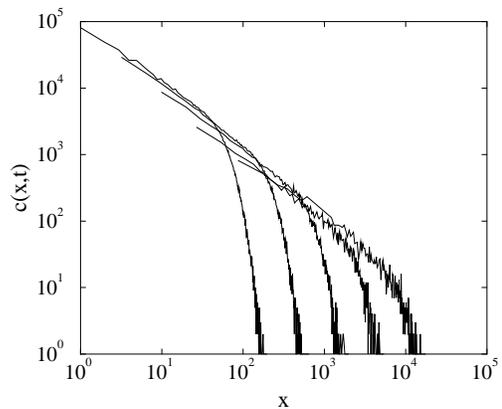}
\caption{The power-law wealth distribution $c(x)$ of greedy multiplicative
  exchange with $\alpha=0.5$ on a double logarithmic scale for times
  $t=1.5^n$, with $n=7$, 10, 13, and 16.}
\label{GE}
\end{center}
\end{figure}

When only the richer agent gains in an exchange, the master equation is now
\begin{equation}
\begin{split}
\label{mge}
\frac{\partial c(x)}{\partial t}= -c(x) +
\tfrac{1}{1-\alpha}\,c\big(\tfrac{x}{1-\alpha}\big)
\int_{x/(1-\alpha)}^\infty dy\, c(y) ~~~~~~\\
~~~~~+\tfrac{1}{\alpha}\int_{x/(1+\alpha)}^x 
dy\, c(y)\,c\big(\tfrac{x-y}{\alpha}\big)\,.
\end{split}
\end{equation}
Numerically, we find that the resulting wealth distribution is a power law
(Fig.~\ref{GE}), with most of the population impoverished.  Pervasive
impoverishment arises because greedy exchange causes the poor to become
poorer and the rich to become richer, but wealth conservation forces there to
be many more poor than rich agents.  In the long-time limit, a small fraction
of the population possesses most of the wealth.

An exact formal solution to Eq.~\eqref{mge} is~\cite{IKR98}
\begin{equation}
\label{exact}
c(x,t)=\frac{A}{x t}, \quad{\rm with}\quad A=-\frac{1}{\ln(1-\alpha)}~.
\end{equation}
This distribution is pathological, however, because positive the moments
$M_n(t)$ of this distribution are divergent.  Thus Eq.~(\ref{exact}) can only
apply within an intermediate scaling regime $x_1(t)<x<x_2(t)$, a restriction
that leads to finiteness of all the moments.  To determine this scaling
region, we use Eq.~(\ref{exact}) to compute the moments and obtain:
\begin{eqnarray}
\label{m01}
M_0(t)&\sim & \int_{x_1}^{x_2} dx\,c(x,t)\sim \frac{A\ln (x_2/x_1)}{t}~,
\nonumber\\
M_1(t)&\sim & \int_{x_1}^{x_2} dx\,x\,c(x,t)\sim \frac{Ax_2}{t}~.
\end{eqnarray}
Since $M_0=1$ and $M_1$ are constants, we infer that $x_1(t)\sim
e^{-t/A}=(1-\alpha)^t$ and $x_2(t)\propto t$.  These cutoffs correspond to
the wealth of the poorest and richest agent, respectively, in the population.
It is only within these ranges that the wealth distribution is a power law,
as shown in Fig.~\ref{GE}.

\section{ DISCUSSION}

Asset exchange represents a parsimonious mechanism for the gain and loss of
individual wealth in an economically active population.  In spite of the
obvious shortcomings of considering only this single factor among the myriad
of influences on individual wealth, asset exchange models lead to a rich
array of wealth distributions.  For additive asset exchange, the wealth
distribution can be explicitly derived for a variety of microscopic exchange
rules.  For greedy multiplicative exchange, where the richer agent always
gains in an interaction, a scaling-based approach indicates that the wealth
distribution has an evolving power law form, $c(x,t)\propto 1/(xt)$.

Power-law distributions occur in the high-wealth tail of the wealth
distribution in various economies, with the associated exponent in the range
of 1.6--2.2 (see Refs.~\cite{ls4,CC07}).  As alluded to in the introduction, a
variety of stochastic models, where agent undergoes an independent stochastic
process, have also been invoked to argue for this power
law~\cite{cha,man,ls2,sor,tak}.  In contrast, greedy multiplicative exchange
is based on a combination of stochasticity and microscopic interactions
between agents.  There are many directions in which asset exchange models
have been extended to make them more realistic; recent work along these
directions can be found in
Ref.~~\cite{DXW06,LKY06,MGI07,C09,SG09,PHC10,SMK10}.  Specific examples of
such additional elements include the incorporation of the saving of
assets~\cite{CC00,CCM04,CC07}, speculative trading~\cite{CPT05}, and other
forms of wealth redistribution.  The notion of exchange of assets has also
been applied to construct a migration model for the distribution of city
sizes~\cite{LR02}.  It should prove interesting to examine the role of such
redistribution mechanisms in the ideologically-free setting of statistical
physics modeling.  The underlying assumption of conserved assets in an
exchange neglects the possibility of wealth growth because of the
exploitation of a natural resource, technological developments, or by both
agents benefiting in exchanges.  These are issues that appear ripe for
further development.

\smallskip We thank Slava Ispolatov for his initial collaboration on this
project.

\end{document}